\documentclass[aps,superscriptaddress]{revtex4}
\usepackage{epsfig,graphics,amssymb,amsmath,color,bm,tabulary,booktabs}
\usepackage{subfigure}
\def\Ca {{\rm Ca}}
\def\Caa {\widetilde{\rm Ca}}
\usepackage{units}

\begin{document}

\title{Crawling Beneath the Free Surface:  Water Snail Locomotion}
\author{Sungyon Lee}
\affiliation{
Hatsopoulos Microfluids Laboratory, Department of Mechanical Engineering,\\
Massachusetts Institute of Technology, 77 Massachusetts Avenue, Cambridge, Massachusetts 02139
}%
\author{John W. M. Bush}
\affiliation{
Department of Mathematics, Massachusetts Institute of Technology,\\ 77 Massachusetts Avenue, Cambridge, Massachusetts 02139
}%
\author{A. E. Hosoi}
\affiliation{
Hatsopoulos Microfluids Laboratory, Department of Mechanical Engineering,\\
Massachusetts Institute of Technology, 77 Massachusetts Avenue, Cambridge, Massachusetts 02139
}%
\author{Eric Lauga\footnote{Corresponding author. Email: \texttt{elauga@ucsd.edu}}}
\affiliation{
Department of Mechanical and Aerospace Engineering, University of California San Diego,\\ 9500 Gilman Drive, Cambridge, La Jolla CA 92093-0411
}%

\date{\today}

\begin{abstract}
Land snails move via adhesive locomotion. Through muscular contraction and expansion of their foot, they transmit waves of shear stress through a thin layer of mucus onto a solid substrate.  Since a free surface cannot support shear stress, adhesive locomotion is not a viable propulsion mechanism for water snails that travel inverted beneath the free surface.  Nevertheless, the motion of the freshwater snail, \textit{Sorbeoconcha physidae}, is reminiscent of that of its terrestrial counterparts, being generated by the undulation of the snail foot that is separated from the free surface by a thin layer of mucus.  Here, a lubrication model is used to describe the mucus flow in the limit of small amplitude interfacial deformations.  By assuming the shape of the snail foot to be a traveling sine wave and the mucus to be Newtonian, an evolution equation for the interface shape is obtained and the resulting propulsive force on the snail is calculated.  This propulsive force is found to be non-zero for moderate values of Capillary number but vanishes in the limits of high and low Capillary numbers.  Physically, this force arises because the snail's foot deforms the free surface, thereby generating curvature pressures and lubrication flows inside the mucus layer that couple to the topography of the foot.
\end{abstract}

\maketitle

\section{Introduction}

Engineers often look to nature's wide variety of locomotion strategies to inspire new inventions and robotic devices \cite{Birch2001,Chan2005,Altendorfer2001,2006SPIE.6230E..86P}.  More generally, scientists across all disciplines are interested in understanding the physical mechanisms behind different styles of biolocomotion.  The mechanism of terrestrial snail locomotion has been investigated and elucidated over the last couple of decades; conversely, the propulsion of water snails that crawl beneath a free surface has yet to be considered.  The purpose of this paper is to propose a propulsive mechanism for  water snails.  

Gastropod locomotion has been of scientific interest for more than a century \cite{Simroth1879,Vl`es1907,Parker1911}. Three distinct modes of locomotion have been examined: ciliary motion, pedal waves, and swimming.  Ciliary locomotion, characterized by the beating of large arrays of cilia on the animal's foot, is usually distinguished from pedal waves by indirect means, such as lack of visible muscle undulation, uniform adherence of the foot to the substrate, and a uniform gliding of the snail body \cite{Audesirk1985}.  This particular type of locomotion is mostly employed by various marine and freshwater snails.

A significant effort has gone towards understanding pedal wave locomotion by terrestrial snails.  Lissmann \cite{Lissman1945,Lissman1945a} was a pioneer in constructing a mechanistic model of the snail foot undergoing such locomotion.  In Ref.~\cite{Lissman1945}, he studied three species of terrestrial snails (\textit{Helix}, \textit{Haliotis}, and \textit{Pomatias}), all of  which use waves of contraction that propagate in the direction of their motion (``direct waves'').  Waves traveling in the opposite direction (``retrograde waves'') were examined by Jones and Trueman \cite{Jones1970}.  A vital insight was later provided by Denny \cite{Denny1980}, who  turned the focus of study from the snail foot to the properties of the pedal mucus.  Pedal mucus has a finite yield stress that allows it to act as an adhesive under small strains and to flow like a viscous liquid beyond its yield point.  Thus, the snail is able to create regions of flow in the mucus by locally shearing it while the rest of the mucus is effectively glued to the solid substrate; these regions (or shear waves) propagate along the length of the foot, enabling the snail to move.  Denny used the nonlinear nature of the mucus to rationalize the locomotion of \textit{Ariolimax columbianus}, a terrestrial slug \cite{Denny1981}. 
Recently, it was found that mucus with shear-thinning properties results in energetically favorable locomotion \cite{Lauga2006}, which is well supported by the experimental studies of the mucus properties.  A detailed investigation of the rheology of mucus was conducted by Ewoldt {\it et. al} \cite{Ewoldt2007}, who also tested different synthetic slimes.  The versatility of snail locomotion 
has also  inspired Chan {\it et. al} \cite{Chan2005}  to build a robotic snail, the first mechanical device to utilize the nonlinear properties of this synthetic mucus. 

Terrestrial snails that employ adhesive locomotion are only a fraction of species in the class of gastropods.  Traditionally, gastropods have been divided into four subclasses: prosobranchia, opisthobranchia, gymnomorpha, and pulmonata, the latter of which consists primarily of terrestrial snails. Many species have not been thoroughly investigated but exhibit interesting locomotive behavior.  For instance, opisthobranchs, that have reduced or absent shells, swim \cite{Farmer1970} or burrow \cite{Aspey1976a,Aspey1976b}.  A still more puzzling mode of locomotion is observed in certain species of water snails.   

In 1910, Brocher \cite{Brocher1910} remarked on water snails that can swim inverted beneath the water surface; since then, other qualitative descriptions have been reported.  Milnes and Milnes \cite{Milne1948} observed the foot of a pond snail ``pulsing with slow waves of movement from aft to fore along its length," suggesting that direct waves are employed for propulsion.  The presence of a trail of snail mucus was also reported.  Goldacre \cite{Goldacre1949} measured the surface tension of this thin trailing film to be approximately $\unit[10]  {dynes/cm}$; he also remarked that the creature was ``grasping the film" as evidenced by the film's being pushed sideways as the snail advanced.  Deliagina and Orlovsky \cite{Deliagina1990} made similar observations while studying feeding patterns of \textit{Planorbis corneus}.  This particular freshwater snail crawls at about $\unit[15]  {mm/s}$, a speed comparable to that on land, while the cilia apparent on the organism's sole ``beat intensely."  Cilia-aided crawling beneath a free surface was observed on marine snails as early as 1919.  Copeland \cite{Copeland1919} concluded that the locomotion of \textit{Alectrion trivittata}, that crawls upside down on the surface, relied solely on the ciliary action.  He conducted a similar study on \textit{Polinices duplicata} and \textit{Polinices heros}, both of which were observed to use both cilia and muscle contraction for locomotion on hard surfaces \cite{Copeland1922}.  Only ciliary motion was employed by the young \textit{Polinices heros} when crawling inverted beneath the surface.  
 
It is therefore clear that freshwater and some marine snails have the striking ability to move beneath a surface that is unable to sustain shear stresses.  In this paper, we  attempt a first quantitative rationalization of these observations. We use  a simplified  model based on the lubrication approximation to show that a free-moving organism located underneath a free surface can move using traveling-wave-like deformation of its foot. We first present our  observations of the propulsion of the freshwater snail, \textit{Sorbeoconcha physidae} in \S\ref{sec:pre_obs}.  We introduce  our model based on the lubrication approximation  in \S\ref{sec:analysis}, and present solutions for small-amplitude motion of the foot. The physical picture for the generation of propulsive forces is discussed in \S\ref{sec:diss}, together with the main conclusions and a summary of the simplifying assumptions used in our analysis.

\section{Observations}
\label{sec:pre_obs}

\begin{figure}[t!]
	\centering
	\includegraphics[width=9cm]{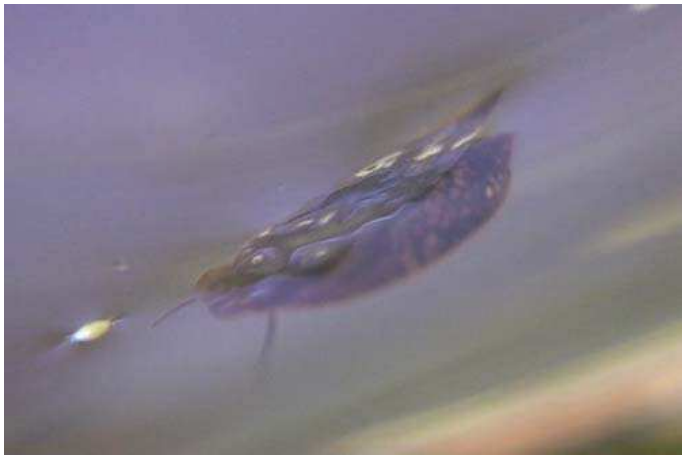}
	\caption{Snail (\textit{Sorbeoconcha physidae}) crawling smoothly underneath the water surface while the surface deforms.  Note the surface deflection associated with the undulatory waves propagating from nose to tail along its foot.  Photo courtesy of David Hu and Brian Chan  (MIT).}
	\label{fig:snailboil}
\end{figure}

\begin{figure}[t!]
	\centering
	\includegraphics[width=9cm]{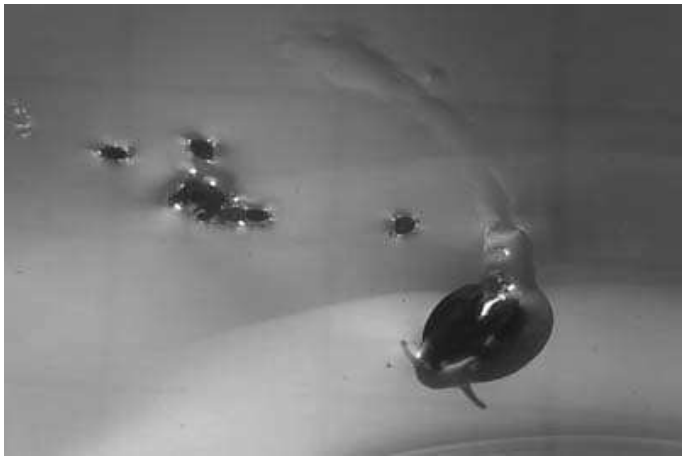}
	\caption{A trail of mucus behind the snail crawling upside down beneath the free surface.  Photo courtesy of David Hu and Brian Chan (MIT).}
	\label{fig:mucus}
\end{figure}

Several common freshwater snails, \textit{Sorbeoconcha physidae}, were collected from Fresh Pond, Massachusetts.   About $\unit[1]  {cm}$ in length, this particular snail can crawl beneath the water surface at speeds as high as $\unit[0.2]  {cm/s}$, comparable to its speed on solid substrates, and perform a $180^\circ$ turn in 3 seconds.  It is rendered neutrally buoyant by trapping air in its shell.  

The undulation of the snail foot causes surface deformations with a characteristic wavelength of $\unit[1]  {mm}$ and amplitude of $\unit[0.2-0.3]  {mm}$ (see Figure \ref{fig:snailboil}).  This deformation appears to travel in the opposite direction of the snail motion, suggesting the generation of retrograde waves, contrary to observations of Milnes and Milnes \cite{Milne1948}.  Another notable feature of water snail propulsion is the presence of a trail of mucus (see Figure \ref{fig:mucus}). For land snails, this mucus layer is typically $\unit[10-20] {\mu m}$ in thickness  \cite{Ewoldt2007}; as with land snails its rheological characteristics may also play a significant role in underwater locomotion.  Since these water snails are also able to crawl on solid substrates, one might venture that their mucus properties do not differ too greatly from those of land snails.  

\section{Model}
\label{sec:analysis}

\subsection{Assumptions}
The crawling of water snails beneath the free surface has four distinct physical features: a free surface with finite surface tension, $\sigma$, a layer of (presumably) non-Newtonian mucus, coupled deformations of the foot and the surface, and a matching of the flow inside the mucus to that around the snail.  To isolate the critical influence of the first feature, we consider in this paper a simplified model system characterized by a Newtonian mucus layer and small deformations of the foot and the interface, with hopes of providing physical insight into the propulsion mechanism.

\subsection{General equations}

Choosing a characteristic velocity $U\sim\unit[1]  {cm/s}$, a mucus thickness $H\sim\unit[20]  {\mu m}$, and a (post-yield) mucus viscosity $\nu\sim\unit[10^{-2}] {m^2/s}$ \cite{Ewoldt2007} suggests a Reynolds number of the flow within the mucus layer to be $Re\equiv{UH}/{\nu} \sim 10^{-5}$.  Thus, we neglect inertia and start with incompressible Stokes equations:
		\refstepcounter{equation}
		$$
		\nabla\cdot\mathbf{v} = 0,\qquad
		\nabla\cdot\mathbf{\Pi} = 0, 
		\eqno{(\theequation{\mathit{a}, \mathit{b}})}\label{eq:govern_eq}
		$$
		where $\mathbf{v}$ is the velocity field inside the mucus and $\mathbf{\Pi}$ the stress tensor.
Normal and tangential stress boundary conditions at the surface may be expressed as 	
		\refstepcounter{equation}
		$$
		\mathbf{n}\cdot\mathbf{\Pi}\cdot\mathbf{n} = \sigma\kappa, \qquad
		\mathbf{t}\cdot\mathbf{\Pi}\cdot\mathbf{n} = 0,
		\eqno{(\theequation{\mathit{a}, \mathit{b}})}\label{eq:freesurf}
		$$
where $\mathbf{n}$ and $\mathbf{t}$ denote, respectively, unit vectors normal (outward) and tangent to the free surface, and  $\kappa = -\nabla\cdot\mathbf{n}$, denotes the curvature of free surface.  We limit our attention to the two-dimensional case, for which the interface shape is given by $\widehat{y} = \widehat{h}(\widehat{x})$ (see Figure~\ref{fig:retro_diag}) and $\kappa$ may be expressed as 
${{\widehat{h}_{\widehat{x}\widehat{x}}}}/{(1+\widehat{h}_{\widehat{x}}^2)^{3/ 2}}$, where the ``hat'' notation denotes dimensional variables.  The mucus is assumed to be a Newtonian fluid; thus, the stress tensor is given by
		\begin{equation}\label{eq:New_stress_tensor}
			\mathbf{\Pi} = -\widehat{p}\mathbf{I} + 2\mu\bm{e}, 
		\end{equation}
where $\bm{e} \equiv{\frac{1}{2}}\{\left(\nabla\mathbf{v}\right) + \left(\nabla\mathbf{v}\right)^T\}$ is the rate-of-strain tensor.

In the frame moving with the snail, we assume that the gastropod foot undergoes periodic deformations in the form of a traveling wave moving at a speed $\widehat{V}_w$. Our approach is as follows: given the shape of the foot, we solve for the shape of the liquid-air interface together with the velocity field in the mucus layer, and then calculate the resulting propulsive force on the snail.

\subsection{Lubrication Analysis}\label{sec:lub}		

\begin{figure}[t!]
	\centering
	\includegraphics[width=9cm]{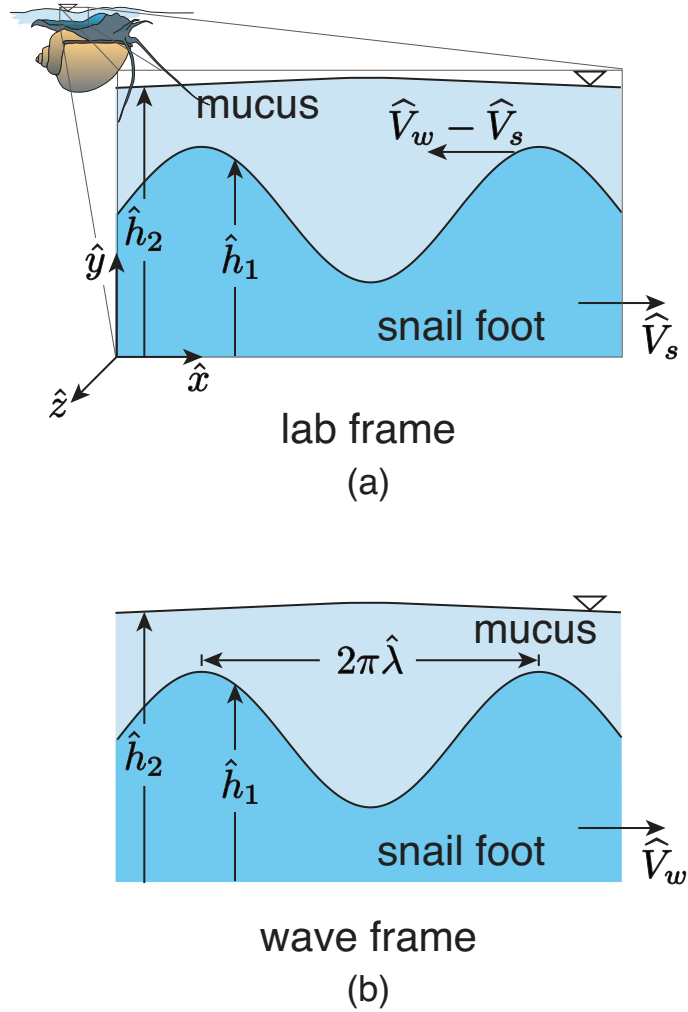}
	\caption{A close-up view of the mucus and the snail foot undergoing a simple sinusoidal deformation of  wavelength $2\pi \widehat{\lambda}$.   The prescribed shape of the snail foot is denoted as  $\widehat{h}_1$; the resultant shape of the free surface, $\widehat{h}_2$, is to be solved for.  The known constant speed of the wave, $\widehat{V}_w$, is set relative to the snail that is translating with an unknown speed, $\widehat{V}_s$.  In laboratory frame (a), the wave is moving in the negative $\widehat{x}$-direction with $\widehat{V}_w-\widehat{V}_s$ while the snail is moving in the positive $\widehat{x}$-direction with $\widehat{V}_s$.  In the frame moving with the wave (b), the snail body appears to move in the positive $\widehat{x}$-direction with $\widehat{V}_w$.}
	\label{fig:retro_diag}
\end{figure}

To eliminate temporal variation, we consider the frame of reference moving at the wave speed $\widehat{V}_w$ relative to the snail (Fig.~\ref{fig:retro_diag}).  We define $\widehat{V}_w$ and the (unknown) snail speed, $\widehat{V}_s$, to be positive when the snail moves in the positive $\widehat{x}$-direction while the wave travels in the opposite direction.  
 
The thickness of the mucus (tens of microns) is observed to be small relative to a typical wavelength of foot deformation (millimeters).  Thus, we apply the lubrication approximation 
\cite{Reynolds1886,Myers1998,Oron1997,Leal2007} and reduce the governing equations based on ${\widehat{H}/\widehat{\lambda}} \equiv a \ll 1$, where $\widehat{H}$ is the characteristic thickness of the mucus film, and $\widehat{\lambda}$ the wavelength divided by $2\pi$.   More specifically, terms of order $a$ or higher are discarded in the equations of motion.	
The governing equations and boundary conditions (Eq.~\ref{eq:govern_eq} - Eq.~\ref{eq:New_stress_tensor}) are non-dimensionalized using the following set of characteristic scales:
	\begin{subequations}
		\begin{align}
			\widehat{x} &= \widehat{\lambda}x,\label{eq:scale_x}\\
			\widehat{y} &= \widehat{H}y,\label{eq:scale_y}\\
			(\widehat{u}, \widehat{v}) &= \widehat{V}_w(u, av) ,\label{eq:scale_u}\\
			\widehat{V_s} &= \widehat{V}_w V_s,\label{eq:scale_V_s}\\
			\widehat{p} &= \frac{\mu\widehat{\lambda}\widehat{V}_w}{\widehat{H}^2}p. \label{eq:scale_p}
		\end{align}
	\end{subequations}		
Based on standard lubrication theory, the equations of motion are reduced to the following:
	\begin{subequations}
		\begin{align}
			0 &= \frac{\partial u}{\partial x} + \frac{\partial v}{\partial y},\label{eq:lubric_1}\\
			0 &= -\frac{\partial p}{\partial x} + \frac{\partial^2u}{\partial y^2},\label{eq:lubric_2}\\
			0 &= \frac{\partial p}{\partial y},\label{eq:lubric_3}
		\end{align}
	\end{subequations}
while the boundary conditions may be expressed as
	\begin{subequations}
		\begin{align}
			u &= 1, \quad\textrm{at}\quad y = h_1,\label{eq:lubric_4}\\
			v &= 0, \quad\textrm{at}\quad y = h_1,\label{eq:lubric_5}\\
  			0 &= \frac{\partial u}{\partial y}, \quad\textrm{at}\quad y = h_2,\label{eq:lubric_6}\\
			{\frac{{a^3}}{ \Ca}}h_{2,xx} &= -p, \quad\textrm{at}\quad y = h_2,\label{eq:lubric_7}
		\end{align}
	\end{subequations}
where $x$ and $y$ are horizontal and vertical coordinates of the system while $z$ points out of the page, and $u$ and $v$ are velocity components in $x$- and $y$-directions, respectively.  The shape of the snail foot is prescribed by $h_1(x)$ while $h_2(x)$ denotes the unknown free surface shape.  Note that $\Ca \equiv{\mu\widehat{V}_w}/{\sigma}$ is not a Capillary number in the traditional sense since $\widehat{V}_w$ is not necessarily the characteristic speed of the flow in the lubrication layer.  
In order to allow surface tension effects to remain relevant in the current problem, the curvature term in Eq.~\eqref{eq:lubric_7} is retained despite being multiplied by $a^3$; this is a standard practice in thin film problems with surface tension (see, e.g., Goodwin and Homsy \cite{Goodwin1991}). 

For convenience, we define a modified Capillary number, $\Caa \equiv {\mu\widehat{V}_w/ a^3\sigma} = {\Ca/ a^3}$, so that the normal stress condition becomes
	\begin{equation}\label{eq:mod_Ca}
		{\frac{1}{ \Caa}}h_{2,xx} = -p, \quad\textrm{at}\quad y = h_2.
	\end{equation}
Then by integrating Eq.~\eqref{eq:lubric_2} twice with respect to $y$, and applying necessary boundary conditions, we obtain an expression for the velocity field in the mucus layer,
	 	\begin{equation}\label{eq:New_vel}
	 		u(x,y) = {\frac{1}{ \Caa}}h_{2,xxx}\left(h_2y - {\frac{1}{2}}y^2 + {\frac{1}{2}}h_1^2 - h_2h_1\right) + 1. 
		\end{equation}
The resulting volume flux through the layer,			
	\begin{equation}\label{eq:New_flux}
		 	Q = h_2 - h_1 + {\frac{1}{ \Caa}}h_{2,xxx}\left[{\frac{1}{3}}\left(h_2 - h_1\right)^2\left(
			\frac{1}{2}h_1 + h_2\right) + h_1\left(h_2 - h_1\right)\left({\frac{1}{2}}h_1 - h_2\right)\right],
		\end{equation}
is constant since the mucus thickness does not vary with time in this moving reference frame.

In order to obtain the motion of the snail, it is necessary to consider the forces acting on the organism.  Since its motion occurs at low Reynolds numbers, the snail is force-free; hence, the forces from the (internal) mucus flow and those from the external flow around the body must sum to zero:
	\begin{equation}\label{eq:force_free_general}
		\mathbf{F}_\text{int} + \mathbf{F}_\text{ext} = \mathbf{0}.
	\end{equation}
More specifically, $\mathbf{F}_\text{ext}$ is equal to $-\widehat{F}_\text{drag}\mathbf{e_x}$, where $\widehat{F}_\text{drag}$ is the magnitude of the drag force from the external flow; $\mathbf{F}_\text{int}$ is the traction caused by the flow in the mucus on the foot of the snail and can be expressed as the integral of $\mathbf{\Pi}\cdot\mathbf{n}_f$, where $\mathbf{n}_f$ is outward normal to the foot of the snail.  In the lubrication limit, Eq.~\eqref{eq:force_free_general} reduces to
	\begin{equation}\label{eq:New_force_bal}
		\widehat{w}\left(\frac{\mu\widehat{V}_w}{a}\right)\int_0^{2n\pi} \left[ p{dh_1\over dx} + \frac{du}{dy}\bigg|_{y=h_1}\right]dx = \widehat{F}_\text{drag},
	\end{equation}
which is a scalar equation representing force-balance in the $x$-direction.  Here $n$ is the number of waves generated by the foot and $\widehat{w}$ is the width of the foot in the $z$-direction.  Physically, the left hand side of  Eq.~\eqref{eq:New_force_bal} is the propulsive force that arises from the internal flow of the mucus and balances the drag from the external flow.  Note that Eqs.~\eqref{eq:force_free_general}-\eqref{eq:New_force_bal} implicitly neglect the overlap regions between the internal mucus flow and the external flow around the organism; we will derive in ~\S\ref{sec:matching} the asymptotic limit in which this is a valid assumption.

\subsection{Solution for small-amplitude motion}\label{sec:result}
In order to solve the model problem, we consider the following limit for foot deformations. If $\Delta\widehat{H}$ denotes the typical amplitude of the foot deformation, we define 
$\varepsilon\equiv{\Delta\widehat{H}/\widehat{\lambda}}$, and assume it to be small. Note that $\varepsilon$  is a parameter independent of the geometrical aspect ratio, $a$, as can be seen by considering the case where  $\varepsilon=0$;  in this limit, the dimensionless parameter, $\varepsilon$, is zero when the foot surface is flat, while $a$ remains finite.
We choose the foot shape as
	\begin{equation}
		h_1 = \varepsilon\sin{x},\label{h1_perturb}
	\end{equation}
and solve for the associated layer profile, $h_2$, order by order as
	\begin{equation}
		h_2 = 1 + \varepsilon h_2^{(1)} + \varepsilon^2 h_2^{(2)} + O(\varepsilon^3).
		\label{h2_perturb}
	\end{equation}
The resulting expression for the flux is given by
	\begin{align}\label{eq:New_flux_perturb}
		\nonumber&\frac{Q}{1+\varepsilon\left(h^{(1)}_2 - \sin{x}\right) + \varepsilon^2 h_2^{(2)}} = 1 +\\
		\nonumber&{(\varepsilon h^{(1)}_{2,xxx} + \varepsilon^2 h^{(2)}_{2,xxx})\over \Caa}\left[{1\over 3}\left(1+\varepsilon\left(h^{(1)}_2 - \sin{x}\right) + \varepsilon^2 h_2^{(2)}\right)^2\left(1+\varepsilon\left(h^{(1)}_2 + {1\over2}\sin{x}\right) + \varepsilon^2 h_2^{(2)}\right)\right] - \\
		&{(\varepsilon h^{(1)}_{2,xxx} + \varepsilon^2 h^{(2)}_{2,xxx})\over \Caa}\left[\varepsilon\sin{x}\left(1+\varepsilon\left(h^{(1)}_2 - \sin{x}\right) + \varepsilon^2 h_2^{(2)}\right)\left(1 + \varepsilon\left(h^{(1)}_2 - {1\over2}\sin{x}\right) + \varepsilon^2 h_2^{(2)}\right)\right],
	\end{align}
where $Q = Q^{(0)} + \varepsilon Q^{(1)} + \varepsilon^2 Q^{(2)} + O( \varepsilon^3)$.  Collecting terms of the same order, the leading order $O(1)$ simply states that $Q^{(0)} = 1$, and then at order $O(\varepsilon)$ we obtain
	\begin{equation}\label{eq:Q_perturb1}
		Q^{(1)} = h^{(1)}_2 + {1\over3\Caa}h^{(1)}_{2,xxx} - \sin{x}.
	\end{equation}
This third order linear ODE has an analytic solution given by
	\begin{align}\label{eq:h2_sol_full}
		\nonumber h^{(1)}_2 &= Q^{(1)} + A_1\exp\left(-{x\over C^{1/3}}\right) + A_2\exp\left({x\over 2C^{1/3}}\right)\cos{\left(\sqrt{3}x\over 2C^{1/3}\right)} \\
		&+ A_3\exp\left({x\over 2C^{1/3}}\right)\sin{\left(\sqrt{3}x\over 2C^{1/3}\right)} + \frac{C\cos{x} + \sin{x}}{C^2 + 1}, 
	\end{align}
where $C \equiv {1\over3\Caa}$, and $A_1$, $A_2$, and $A_3$ are unknown constants.  For convenience, $Q^{(1)}$ is set to zero by arbitrarily setting $Q \equiv 1$.  Note that $Q$ corresponds to the rate of mucus production by the snail.

\subsection{Boundary conditions}
\begin{figure}[t!]
	\centering
	\includegraphics[width=7cm]{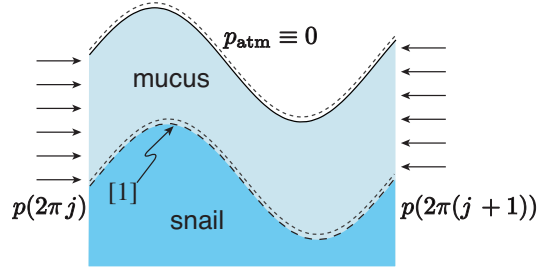}
	\caption{Free body diagram of a perfectly periodic mucus layer over one wavelength between nodes $j$ and $j+1$.  Pressures at these nodes, $p(2\pi j)$ and $p(2\pi(j+1))$, as well as the heights, $h_2(2\pi j)$ and $h_2(2\pi(j+1))$, are equal by the periodic boundary conditions.  Above the mucus layer is open to atmosphere with $p_\text{atm}$ set to zero.}
	\label{fig:periodic_CV}
\end{figure}

The real challenge lies in identifying the three independent boundary conditions required to solve for $A_1$, $A_2$, and $A_3$.  As a logical starting point, we proceed by applying periodic boundary conditions over each wavelength:
	\begin{subequations}
	\begin{align}
		\label{eq:per_bc1}h_{2,x}(2\pi j) &= h_{2,x}(2\pi (j+1)),\\
		\label{eq:per_bc2}h_{2,xx}(2\pi j) &= h_{2,xx}(2\pi (j+1)),\\
		\label{eq:per_bc3}h_{2,xxx}(2\pi j) &= h_{2,xxx}(2\pi (j+1)),
	\end{align}
	\end{subequations}
where $j$ ranges from $0$ to $n-1$.  In this limit, $A_1$, $A_2$, and $A_3$ vanish, and these boundary conditions yield no motion of the snail, regardless of the value of $\Caa$.  Conducting a force balance on the mucus layer over one wavelength (as shown in Fig.~\ref{fig:periodic_CV}) offers a simple explanation for this result: the periodic boundary conditions ensure that the pressure forces acting on the side control surfaces of the mucus layer precisely cancel.  Since the top control surface of the mucus is exposed to ambient air pressure, there can be no net force that acts on the bottom control surface.  Thus, no equal and opposite force acts on the snail foot ([1] in Fig.~\ref{fig:periodic_CV}), suggesting that no net propulsive force is generated under strictly periodic boundary conditions.

Instead, the three  boundary conditions should be selected by the physical constraints on snail locomotion. Since the moving gastropod is force- and torque-free, the first two conditions should naturally be
	\begin{subequations}
	\begin{align}
		\label{eq:bc1}\Sigma F_y &= 0,\\
		\label{eq:bc2}\Sigma\tau &= 0,
		\end{align}
	\end{subequations}
where $F_y$ refers to forces in the $y$-direction while $\tau$ is a torque in the $z$-direction.  
{In all generality, the sum of all forces and torques acting on the snail must vanish at low Reynolds number; the forces and torques from the thin film of mucus must therefore balance the forces and  torques generated by the external flow around the body and those arising from gravity.  For this analysis, we assume the snail does not rotate, and that its shape is sufficiently symmetric that the  external viscous torques and $y$-force vanish. 
We also assume the organism to be neutrally-buoyant and homogeneous, so the forces and torques due to gravity are zero.  As a result, Eqs.~\eqref{eq:bc1} and \eqref{eq:bc2} only require the $y$-force and $z$-torque arising from the thin film to vanish.}  

The final boundary condition arises from consideration of the matching between the internal and the external flow around the organism.  By symmetry, we expect the swimming speed of the snail to be of order $V_s\sim \varepsilon^2$.  Since we are in the Stokes regime, pressure differences across the moving gastropod should scale linearly with the free stream velocity, and occur at order $\varepsilon^2$ as well.  Thus, the pressure difference between the front and back of the snail is zero at $O(\varepsilon)$, the order of our formulation, and the third boundary condition becomes:
	\begin{equation}
		\label{eq:bc3}p(0) = p(2n\pi).
	\end{equation}

These three boundary conditions allow one to obtain a complete expression for $h^{(1)}_2$ and subsequently solve for the dimensionless propulsive force, $F_\text{prop}$ at $O(\varepsilon^2)$:
	\begin{align}\label{eq:force_bal_perturb2}
		\nonumber F_\text{prop} &= \int_0^{2n\pi}\left[ p{dh_1\over dx} + \frac{\partial u}{\partial y}|_{y=h_1}\right]dx\\
		&= 3 C\int_0^{2n\pi} \left[-h^{(1)}_{2,xx}\cos{x} + h^{(1)}_{2,xxx}(h^{(1)}_2 - \sin{x}) + h^{(2)}_{2,xxx}\right]dx.
	\end{align}
Note that the integral of $h^{(2)}_{2,xxx}$ vanishes by the matching pressure boundary condition, which is equivalent to $h_{2,xx}(0) = h_{2,xx}(2n\pi)$.  Referring back to the constants in Eq.~\eqref{eq:h2_sol_full}, $A_1$, in particular, is a non-trival function of $\Caa$.  Hence, unlike the strictly periodic boundary condition case, the expression for $h^{(1)}_2$ now contains a non-periodic function that gives rise to a non-zero propulsive force.

\subsection{Crawling speed}

\begin{figure}[t!]
	\centering
	\includegraphics[width=11cm]{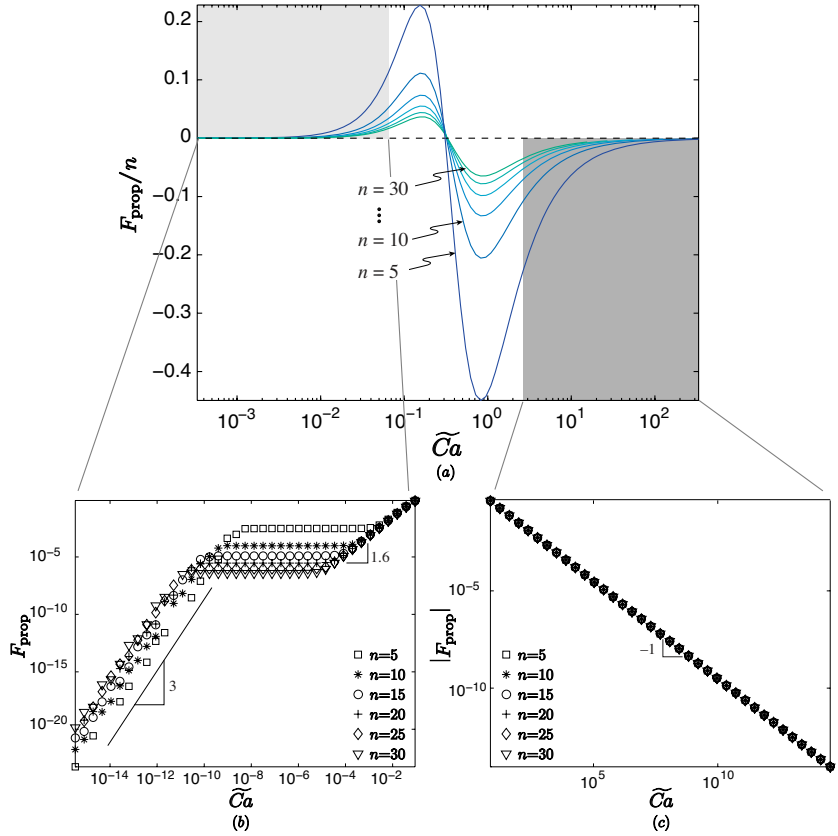}
	\caption{{(a) Dimensionless propulsive force, $F_\text{prop}$, normalized by the number of wavelengths, $n$, as a function of the modified Capillary number, $\Caa \equiv {\mu\widehat{V}_w/ a^3\sigma}$, where the values of $n$ range from 5 to 30 in the increment of 5. In (b) and (c), the absolute value of the  dimensionless force, $F_\text{prop}$, is plotted on a logarithmic scale to show the power-law decay in the limits of $\Caa\rightarrow0$ and $\Caa\rightarrow\infty$ respectively.  The propulsive force  exhibits a $\Caa^{-1}$ decay for large $\Caa$, while it decays as $\Caa^{3}$ for small $\Caa$.}}
	\label{fig:propulsive_plot}
\end{figure}

To balance the thin-film propulsive force, it is necessary to evaluate the external drag $\widehat{F}_\text{drag}$ caused by the motion of the snail.  For simplicity, the snail is modeled as approximately spherical, with radius $\widehat{R}$.  Although approximate, this model yields an order of magnitude approximation for the speed of our model snail.  
Non-dimensionalizing $\widehat{F}_\text{drag}$ by $\mu \widehat{V}_w\widehat{w}$, the right hand side of \eqref{eq:New_force_bal} becomes $F_\text{drag} \approx 6\pi f{V_s/\mu^{*}}$,
where $V_s$ is the snail speed scaled by $ \widehat{V}_w$ and $\mu^{*} \equiv {\mu/\mu_{water}}$, the viscosity ratio of mucus to water.  A correction factor $f$ accounts for the aspherical shape of the snail as well as the influence of the free surface on the drag coefficient; for the present analysis, $f$ will be treated as known for a given crawler.  By combining this scaling with the $O(\varepsilon^2)$ term in the force on the foot generated by the mucus layer, the following expression for $V_s$ is obtained:
	\begin{equation}\label{eq:V_s}
		V_s \approx \frac{\varepsilon^2\mu^{*}}{af}F_\text{prop}(\Caa, n),
	\end{equation}
where $F_\text{prop}$, the total propulsive force function, is plotted in Fig.~\ref{fig:propulsive_plot}(a) for different values of $n$.  The exact formula for $F_\text{prop}$ is not reproduced in this paper for it is long and not informative for the purpose of this analysis, but is straightforward to calculate with symbolic packages.  Note that the width of the snail foot, $\widehat{w}$, is taken to be on the same order as $\widehat{R}$; thus, they drop out of Eq.~\eqref{eq:V_s}.  
	
\subsection{Results}
Fig.~\ref{fig:propulsive_plot} shows that the propulsive force vanishes in the limits of both large and small surface tension.  
In the limit of infinitely large surface tension ($\Caa\rightarrow0$), the interface between air and mucus is undeformable and so is analogous to a flat surface that cannot sustain shear stress.  A snail would simply slip on such surfaces.  The detailed behavior for small values of $\Caa$  is shown in 
Fig.~\ref{fig:propulsive_plot}(b). The dimensionless force follows the power-law decay 
$F_\text{prop} \sim \Caa^{3}$ for decreasing $\Caa$ for all values of $n$.

With zero surface tension ($\Caa\rightarrow\infty$), a pressure difference across the interface cannot be sustained and so cannot drive the flow within the mucus; hence no propulsive force can be generated in this low surface tension limit either. In this case, the force follows for all $n$ the power-law decay  $|F_\text{prop}|\sim \Caa^{-1}$ as shown in Fig.~\ref{fig:propulsive_plot}(c).

Note that the propulsive force goes from positive to negative at a moderate value of $\Caa$ that, in the case of $n=10$, is around $0.3$. Physically, this implies that the snail switches from retrograde waves to direct waves at this critical $\Caa$.  In addition, Fig.~\ref{fig:propulsive_plot}(a) shows that the propulsive force exhibits two distinct maxima for retrograde and direct waves, at values of $\Caa$ corresponding to $0.15$ and $0.8$, respectively.  Since the maximum propulsive force for the direct waves is higher than that for the retrograde, the direct waves may be a faster mode of locomotion for water snails.  This points to a possible biological advantage of  direct over retrograde waves.
\begin{figure}[t!]
	\centering
	\includegraphics[width=6cm]{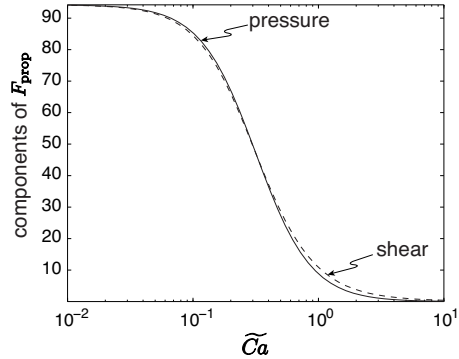}
	\caption{Absolute magnitudes of components of dimensionless propulsive force due to pressure (solid line) and due to shear (dashed line) as a function of $\Caa$ for $n=10$. (note that the shear force is negative).  Hence, the total propulsive force which is the sum of these two forces is non-zero only when there is a difference between the two.}
	\label{fig:px_shear_plot}
\end{figure}

Fig.~\ref{fig:px_shear_plot} quantifies the components of propulsive force due to pressure and shear.  It is important to note that the force due to shear (dashed line) is negative.  In the low $\Caa$ limit, these two components precisely cancel, leading to no motion.  When $\Caa$ is high, they both vanish.  For intermediate values of $\Caa$, a difference in the magnitudes of these forces results in a net propulsive force.  Note the existence of a finite value of $\Caa$ for which the propulsive force reaches zero, which is a surprising result of our model. 

\begin{figure}[t!]
	\centering
	\includegraphics[width=7cm]{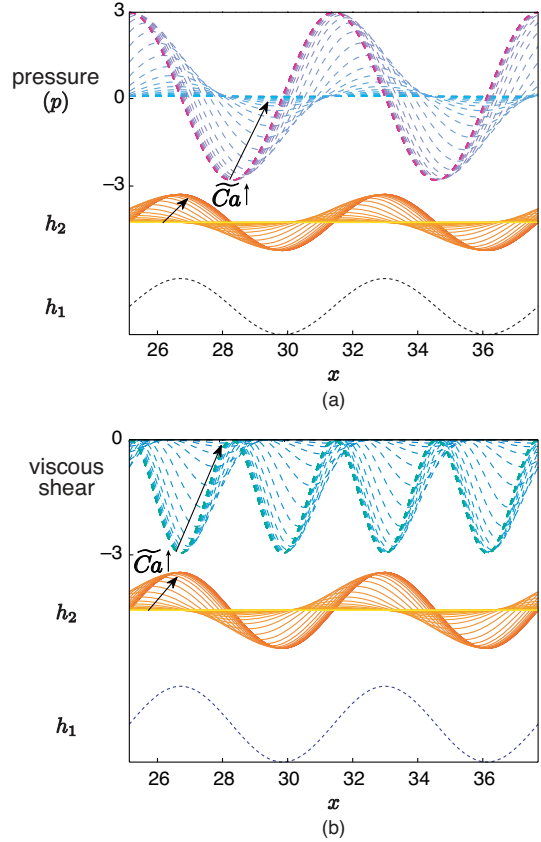}
	\caption{Dimensionless pressure (a, dashed line) and shear stress (b, dashed line) within the mucus over two wavelengths for $n=10$.  The single dotted line in both (a) and (b) is the shape of the foot, $h_1$ while the solid lines describe the shape of the interface, $h_2$ for different values of $\Caa$.  Black arrows indicate the direction of increasing $\Caa$.}
	\label{fig:p_shear_mid}
\end{figure}	 

\begin{figure}[t!]
	\centering
	\includegraphics[width=7cm]{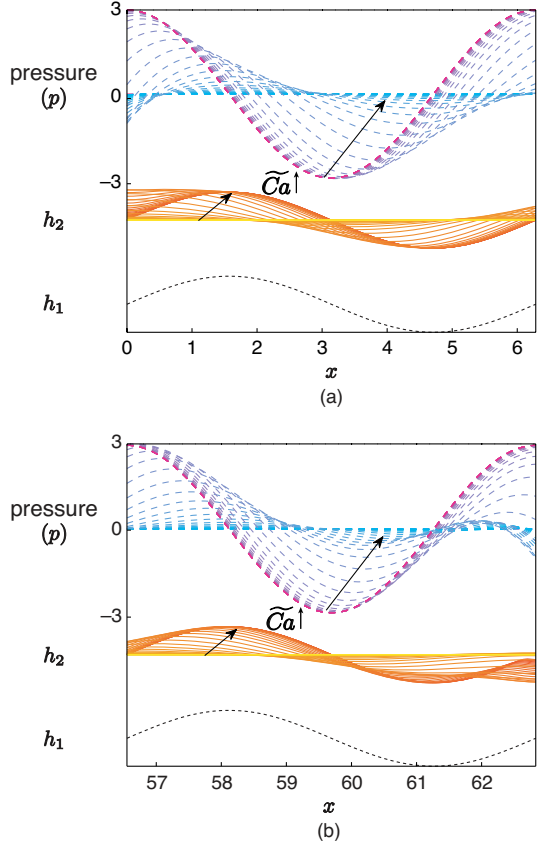}
	\caption{Dimensionless pressure (dashed line) and the interface shape (solid line) in the front (a) and end (b) of the snail for $n=10$.  The dotted line is the shape of the foot, and black arrows are in the direction of decreasing surface tension.}
	\label{fig:p_front_end}
\end{figure}

\begin{figure}[t!]
	\centering
	\includegraphics[width=7cm]{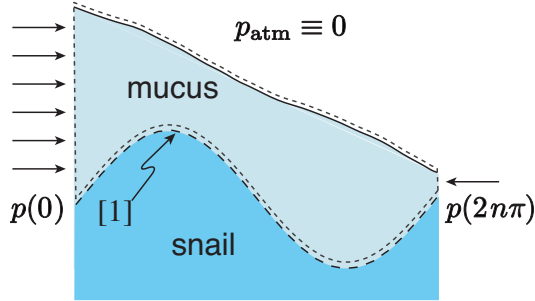}
	\caption{Free body diagram of an asymmetric mucus layer across the foot of the snail.  (For simplicity, $n=1$ in this diagram.)  Pressures at the ends, $p(0)$ and $p(2n\pi)$ are equal by the boundary condition; however, they act over two different mucus thicknesses, resulting in a net pressure force.}
	\label{fig:asym_CV}
\end{figure} 

Going back to the dynamic boundary condition Eq.~\eqref{eq:mod_Ca}, one can calculate the pressure and shear stress distribution to $O(\varepsilon)$ inside the mucus layer.  These are plotted in Figure \ref{fig:p_shear_mid} for $n=10$ along with the shape of the interface, $h^{(1)}_2$.  In the large $\Caa$ limit, the interface shape exactly conforms to the shape of the foot $h_1$; in the small $\Caa$ limit, the interface becomes flat.  At intermediate $\Caa$, there exists an asymmetry in the interface shape, {associated with the exponential term in Eq.~\eqref{eq:h2_sol_full}}, that gives rise to the non-trivial propulsive force.  The interface shape, pressure and shear stresses are plotted for the first wavelength and the last (corresponding to the front and end of the snail) for $n=10$ in Fig.~\ref{fig:p_front_end}.  As shown in this figure, 
the mucus thickness at the ends deviate substantially from extrapolated periodic values creating an asymmetry between the head and the tail. Thus, although the pressures at the ends of the crawler are equal at $O(\varepsilon)$, there exists a net $O(\varepsilon^2)$ pressure force acting on the side control surfaces, owing to the  a $O(\varepsilon)$  difference in thickness of the mucus layer (see Fig.~\ref{fig:asym_CV}).  To balance this net force, there has to be a force acting on the bottom surface of the mucus layer; thus, there exists an equal and opposite force acting on the foot of the snail, corresponding to the propulsive force.  

As suggested by Fig.~\ref{fig:propulsive_plot}, surface tension is the essential ingredient in this mode of locomotion, and the propulsive force vanishes in both limits of asymptotically small (large $\Caa$) and large (small $\Caa$) surface tension.  The snail would therefore have to tune the way it deforms its foot to exploit the property of the fluid-air interface.  As the foot is deformed, it forces a lubrication flow in the mucus layer above and leads to the deformation of the free surface.  The resulting topography of the free surface, constrained by surface tension, is then exploited by the organism to generate a propulsive force. In essence, capillary forces transform the flat free surface into a wavy one capable of sustaining tangential stresses and so enabling propulsion.

\subsection{Matching internal and external flows }\label{sec:matching}

In our analysis, we have neglected the fluid forces on the organism, $\mathbf{F}_\text{match}$, arising from the intermediate matching region between the internal (mucus) and the external flows.  Since the propulsive force of the snail mostly arises from the asymmetric shape of the free surface at the head and tail of the foot, this requires further comment.  

Physically, since we are calculating the internal and external flows separately, both need to be considered.
In order to first estimate the magnitude of $\mathbf{F}_\text{match}$ due to the external flow, we refer to the work by Berdan and Leal \cite{Berdan1982} who studied the motion of a sphere near a deformable fluid-fluid interface.  As an extension of previous work  in which the interface is assumed to be flat \cite{Lee1979,Lee1980}, the work in Ref.~\cite{Berdan1982} considers the limit of small interfacial deformation and its effects on the translating body.  Unlike our current analysis, the velocity of the sphere is not governed by the shape of the free surface but is fixed as $\widehat{U}$.  The small parameter in this paper, $\hat{\epsilon}$, reduces to a Capillary number, $\hat{\epsilon}={\mu\widehat{U}/\sigma}$, when gravitational effects are not included. Berdan and Leal showed that in the case of a sphere moving parallel to the free surface, the deformation of the interface only has a vertical force contribution at $O(\hat{\epsilon})$.  In the current analysis, $\mathbf{F}_\text{int}$ and $\mathbf{F}_\text{ext}$, the forces considered in Eq.~\eqref{eq:force_free_general}, are of $O(\varepsilon^2)$.  Therefore, in order to neglect $\mathbf{F}_\text{match}$ consistently, the following condition has to be satisfied:
	\begin{equation}\label{eq:match_neglect}
		\hat{\epsilon}^2 \ll \varepsilon^2,
	\end{equation}
which requires one to have a look at how $\hat{\epsilon}$ and $\varepsilon$ are defined.  Since $\widehat{U}$ is $\widehat{V}_s$ in our problem, we have $\hat{\epsilon} \sim {\mu\widehat{V}_s/ \sigma}$.  Recalling from ~\S\ref{sec:lub}, the Capillary number, $\Ca$, is defined in terms of the wave velocity, $\widehat{V}_w$.  Because it has been shown that $V_s ={\widehat{V}_s}/{\widehat{V}_w}$ scales as $\varepsilon^2$, $\hat{\epsilon}$ can be expressed as
		\begin{equation}\label{eq:leal_small}
		\hat{\epsilon} \equiv {\mu\widehat{V}_s\over\sigma} \sim \varepsilon^2{\mu\widehat{V}_w\over\sigma} \equiv \varepsilon^2 \Ca.
	\end{equation}
When one replaces $\Ca$ with $a^3\Caa$ and rearranges the terms, the criterion to neglect $\mathbf{F}_\text{match}$ in Eq.~\eqref{eq:match_neglect} reduces to
	\begin{equation}\label{eq:match_con2}
		\varepsilon^2a^6\Caa^2 \ll1.
	\end{equation}  		 

Since $a$ and $\varepsilon$ are both small parameters asymptotically approaching zero in the lubrication analysis in the limit of small deformation amplitude, ${\varepsilon^2a^6\ll 1}$, Eq.~\eqref{eq:match_con2} represents a weak constraint on the validity of neglecting forces from the intermediate region. 

The second matching force to consider is that induced by the internal flow. Since there is, in general, a height difference between the mucus at the front and the back of the snail, the fluid surface will be distorted at either end to match with the flat surface far away. Our work will therefore be valid in the limit where the capillary forces resulting from these distortions can be neglected, corresponding to an asymptotic limit which we now characterize. 

The two relevant length scales to consider for matching  the  distorted fluid interface to the flat free-surface in the far-field 
are the capillary  length, $\ell_c\sim\sqrt{\sigma/\rho g}$ ($\rho$ is the fluid density), and the width of the snail, $\hat w$.The fluid interface will be distorted over a length, $\ell$, into the fluid, where $\ell\sim \min (\ell_c,\hat w)$.
The typical curvature pressure arising from surface distortion will be on the order of $\sim \sigma \delta \hat h/\ell^2$, acting on typical height difference $\delta \hat h$ between the free surface near the snail and the far-field height of the free surface, and therefore contributes to a force on the snail (per unit width) on the order of $\sim \sigma (\delta \hat h)^2/\ell^2$. Since $\delta \hat h\sim \varepsilon \hat H \sim \varepsilon a \hat \lambda  $, the  capillary force  is on the order of 
$\sim  \sigma \varepsilon^2 \hat a^2 \hat \lambda^2/\ell^2$.
This force has to be compared with that arising  from  the external flow, given by $p_{ext} R$
(again, per unit width)  where $p_{ext}$ is the typical magnitude of the pressure outside the organism as it is crawling. 
Since $p_{ext}\sim\mu \hat V_s/R$ we have $p_{ext} R\sim\mu \hat V_s \sim \varepsilon^2\mu \hat V_w$.
The matching condition becomes therefore 
$  \sigma \varepsilon^2 \hat a^2 \hat \lambda^2/\ell^2 \ll \varepsilon^2\mu \hat V_w$,
which is equivalent to
\begin{equation}\label{second}
\hat R ^2/\ell^2 \ll a\, \Caa\, n^2,
\end{equation}
where we have used the estimate $\hat R\sim n\lambda$. The second matching condition,  Eq.~\eqref{second}, requires that the number of wavelengths along the snail's foot, $n$,  be sufficiently large.

\section{Discussion}
\label{sec:diss}

In this paper, we have presented a simplified model of water snail locomotion. The physical picture that emerges is the following: the undulation of the snail foot  causes normal stresses that deform the interface and drive a lubrication flow.  The resulting stress distribution couples to the topography of the snail foot, leading to a propulsive force.  This force vanishes in the limit of $\Caa \to 0$, where the interface is flat, and of $\Caa\rightarrow\infty$, where the topographies of the interface and the snail foot precisely match.  A finite propulsive force is obtained for intermediate values of $\Caa$.  This interplay between the free surface and the snail foot distinguishes water snail locomotion from that of their terrestrial counterparts.  For the latter, the solid substrate on which the snail crawls is fixed; hence, the shape of the snail foot alone determines the pressure and shear stresses generated within the mucus layer.  For water snails, however, the interface is deformed due to the flow created in the mucus by the foot undulation; the interface, in turn, affects the dynamics within the mucus layer, creating pressure and shear stresses that act on the foot.  This nonlinear coupling between the foot geometry, surface tension, and dynamics within the mucus layer makes the water snail locomotion a less straightforward mode of locomotion.

A direct analogy exists between the thin film comprising the mucus layer of water snails and those arising in coating flows; for example, those used in photolithographic processes to fabricate various electronic components. This class of fluids problem has been well studied, both experimentally \cite{Stillwagon1988,Stillwagon1990,Peurrung1993} and theoretically \cite{Stillwagon1988,Stillwagon1990,Pozrikidis1991,Pritchard1992,Schwartz1995,Roy2002,Gaskell2006}, an example of which includes spin coating.  Kalliadasis et al. \cite{Kalliadasis2000} used lubrication theory to show that in the limit of small $\Ca$, the interfacial features become less steep, an effect also captured by our model.  Mazouchi and Homsy \cite{Mazouchi2001} demonstrated that, in the case of large Capillary number, the shape of the free surface nearly follows the topography.  In the context of water snail locomotion, we saw in~\S \ref{sec:result} that the free surface conforms to the shape of the foot in the same limit.  

Our study is only the first step towards a quantitative understanding of  gastropod crawling beneath free surfaces.  It is significant in that we have demonstrated the plausibility of locomotion with the minimal ingredients: Newtonian fluids and small amplitude deformations.  Nevertheless, outstanding issues remain.  In the case of adhesive snail locomotion on land, the non-Newtonian properties of snail mucus, such as a finite yield stress and finite elasticity, play an essential role \cite{Denny1980}.  Non-Newtonian mucus is likewise expected to have a significant effect for water snails. Furthermore, there need to be more systematic observational studies to identify which water snail species exhibit which modes of ``inverted crawling''.  As reported by Copeland \cite{Copeland1919,Copeland1922} and Deliagina and Orlovsky \cite{Deliagina1990}, some species of water snails rely entirely on cilia for propulsion beneath the free surface.  If such ciliary motion results in no free surface deformation, the physical mechanism examined in this paper is of little relevance, and a closer look at the cilia-induced flow is suggested.  Alternatively, the non-Newtonian properties of the mucus may prove to be significant in this case.  Categorizing different species according to their propulsion mechanism of crawling (i.e. cilia versus muscle contraction) and the constitutive properties of their mucus would provide a more complete physical picture of this intriguing form of locomotion.       

 \section*{Acknowledgements}
We thank Brian Chan and David Hu for the pictures. 
We also thank George M. Homsy for helpful discussions.
This work was supported in part by the NSF (grant CTS-0624830).

\bibliography{scribble_ref_eric}
\bibliographystyle{unsrt}

\end{document}